# The EAVI ExG
## Muscle/brain hybrid physiological sensing

Atau Tanaka, David Fierro, Francesco Di Maggio
MSH Paris Nord
[atau.tanaka, david.fierro, francesco.dimaggio]
@mshparisnord.fr

Martin Klang
Rebel Technologies
Barcelona
martin@rebeltech.org

Stephen Whitmarsh
Sorbonne Université
Paris Brain Institute-ICM
Hôpital de la Pitié Salpêtrière
stephen.whitmarsh@icm-institute.org

## ABSTRACT
We present an update on the EAVI physiological interface, an open-source wireless, microcontroller-based hardware design for the acquisition of bioelectrical signals. The system has been updated to process electroencephalogram brain signals in addition to muscle electromyogram. The hardware/firmware system interfaces with host software carrying out feature extraction and signal processing. The demo will show multichannel EMG, and single channel EEG. We call this hybridization "ExG". We will present documentation of the EAVI board used in the lab and on stage, in user studies with neuro-diverse musicians and trained instrumentalists, as well as in performance.

## 1. INTRODUCTION
Recent advances in electronics hardware, signal processing, and information analysis have made physiological computing applications practical and feasible, taking it out of the pure biomedical domain to find applications in human-computer interaction (HCI). Commercial and custom hardware have been used by musicians to create digital musical instruments from physiological computing systems.

However, there is a gap in the market between two extremes. High-end medical grade hardware, while offering excellent signal quality, remains prohibitively expensive and difficult to use outside of clinical settings. Meanwhile the Do-It-Yourself (DIY) movement has produced low-cost alternatives but are built upon low-grade, general purpose amplifiers and converters not specifically tuned to the nature of the physiological signal (often in µ-volts). The hardware design we present here combines a specialized biosignal acquisition chip mated with a general-purpose microcontroller.

The proposed design, the EAVI ExG board, is a DMI system that senses muscle electromyogram (EMG) and brain electroencephalogram (EEG) signals. It is novel in that is combines physiological and audio signals in a single, unified signal processing chain. The device is MIDI and audio class compliant, and designed to integrate into a computer music or modular synthesis performance system with no host-side drivers.

## 2. RELATED WORK
Physiological signals have been detected using analogue electrical circuits and electronics since the 1930s [1]. The EEG has been used for musical performance by composers like Alvin Lucier and David Rosenboom since the 1960s [9,10]. The advent of digital signal processing in the 1980s brought a boon of early DMI work in the field, including the BioMuse, Bodysynth, and Mini BioMuse [4,6,14].

Industrial design and wireless communications in the 2000's enabled technology startups to bring to market novel human interface devices (HID) like the Delsys [5] and Thalmic Labs Myo. While never intended for musical use, these commercial devices were hacked and adopted by musicians and artists [7]. Such communities have been left orphaned as corporate acquisitions have led to products like the Myo being discontinued and unsupported on new computing platforms [3].

The DIY community has used the democratization of hardware development to propose frugal solutions based on common platforms like the Arduino [13,15]. Biomedical companies like Plux have released the Bitalino product line to make creative physiological sensing more accessible [12].

Meanwhile, a new generation of tech startup coming out of academia has focused on filling the void left by the disappearance of the Myo[1] [2], and extend the application to sports medicine[2] [8].

In all these cases, the physiological interface needs to be adapted to musical use. Except for early work like the BioMuse, BodySynth and Mini Biomuse that had MIDI outputs, biomedical devices and general purpose HIDs cannot be considered digital musical instruments as such. We present an affordable wireless physiological interface designed from the ground up to be a musical device.

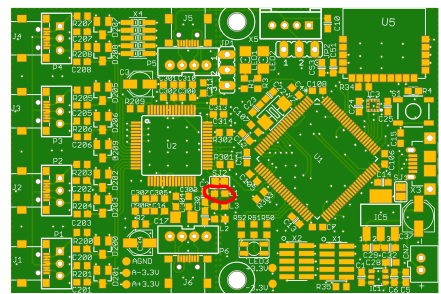

**Figure 1. The EAVI ExG PCB**

---

[1] https://sifilabs.com/

[2] https://lwt3.com/

## 3. HARDWARE DESIGN

The hardware design we present here combines a specialized biosignal acquisition chip mated with a general-purpose microcontroller, extending an earlier prototype [ANON]. It is based on the Texas Instruments ADS129x family a single chip integrated solution for high quality biosignal amplification and digitization[3]. The microcontroller is the STMicroelectronics STM32F427, a Cortex-M4 family microcontroller with floating point unit[4]. In addition to EMG acquisition, the board includes a Kionix KX122 three-axis accelerometer[5]. The TI and Kionix sensing chips communicate with the ST microcontroller over an I2C digital serial bus. The board communicates with the host computer or rest of the music system wirelessly over Bluetooth LE 4.2 using an ST SPBTLE-1S transceiver. The board can also communicate over USB where it registers with the host as a class compliant device.

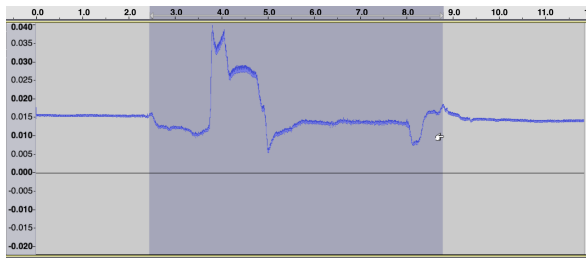

**Figure 2. EMG amplitude envelope from fist clench**

### 3.1 Signal Acquisition

Electrical biosignals are detected through electrodes placed on the skin and connected to a difference amplifier. The signals are then digitized by and Analog to digital converter (ADC) and sent to a microcontroller unit (MCU). Analog and digital signal processing ensures that the resulting data stream is of high resolution, free from aliasing artifacts and noise, and of sufficient bandwidth.

The EMG signal presents challenges in the signal acquisition, including low signal level, large base line drift, and relative proximity of noise frequencies to the signals of interest. The acquired signal may be too noisy to be useful, while filtering may remove salient features. Traditionally this required carefully designed analogue filters preceding digitization. Modern ADCs support a very wide dynamic range, which makes it feasible to digitize the signal directly with minimal analog filtering. The acquired signal includes base line drift and noise that must then be removed through digital signal processing. This allows adjustment adapted to specific use cases.

The high-resolution ADC allows less overall gain, allowing the signal to be tracked across wide baseline drift. Previously drift had to be compensated for in the analogue domain. With 24bit resolution, the dynamic range is large enough to capture small signals. Programmable gain allows the device to be easily tuned to different signal ranges.

We use the TI analogue front end (AFE) to digitize and low pass filter the signal, with decimation filters that provide efficient anti-aliasing. The ADS129x combines precision difference amplifiers with 24-bit sigma-delta ADCs, programmable gain amplifiers, and integrated decimation filters. It is capable of operating with extremely low noise floor, down to less than 5μV RMS. Decimation filters down-sample the input signal from over 200kHz for efficient anti-aliasing. The cut-off frequency is programmable from 5Hz to 1280Hz, easily covering the requirements for EEG, ECG and EMG signals. The output signal is low pass filtered and anti-aliased, but includes base line drift and some amount of powerline noise. The chip has flexible signal routing to improve common mode rejection with integrated Right Leg Drive.

This front end is combined with the micro-controller for digital signal processing at the output signal rate (4 channels at up to 6kHz). The STM32F303 runs at 72Mhz, and is capable of real time audio signal processing. It configures the front end, and connects over a serial interface with a data transceiver.

In addition to the EMG signals, an onboard 3-axis accelerometer provides movement data. The KX122 processes data rates up to 12.9kHz, and user selectable ranges of ±2g, 4g, or 8g acceleration.

The EAVI board captures ExG data at a sample rate of 16 kHz and a resolution of 20-bits.

### 3.2 Signal Pre-processing

Pre-processing consists of a combination of any of the following:
- Sample rate conversion
- High pass or DC filter to remove base line drift
- Data conversion
- Notch filter to remove powerline hum
- Low pass filter to remove EMI noise

Sample rate conversion is used to align the accelerometer data with the sampled EMG signal, and down-sampled before wireless transmission. High pass filtering is required to remove or drastically reduce the base line drift. After this stage, Data conversion can be applied to reduce the bit depth, and dynamic range, without saturation or clipping.

Additional filtering can help reduce power line hum (notch filtering at 50 or 60 Hz) and EMI noise.

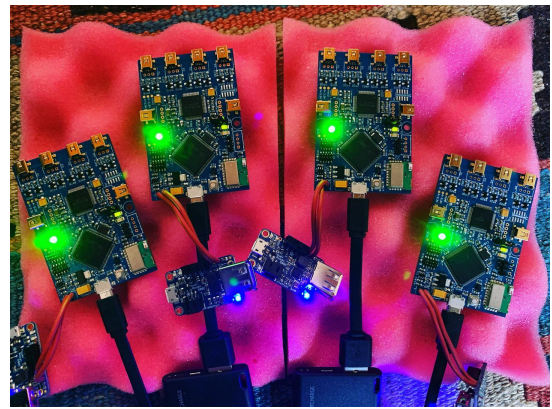

**Figure 3. Four EAVI ExG boards**

### 3.3 Data Transmission

The pre-processed data can be transmitted to a host computer in one of two ways, USB or BLE.

When transmitting over USB, the device implements a class compliant USB Audio interface which allows high rate, high resolution and low jitter / noise multichannel data transmission without requiring device drivers to be installed. However, a USB isolator must be used between the BMI hardware and the host computer, to prevent any risk to the subject through electrical coupling.

With wireless BLE transmission, the bandwidth is restricted which requires sample rate and bit depth reduction in the pre-processing. To allow real-time data transfers to a host device we establish a connection using the BLE MIDI standard profile. This allows interoperability with a wide range of devices: smartphones, tablets and computers. Data is transmitted as MIDI Pitch Bend messages, using one MIDI channel per electrode. This supports up to 16 channels of data, with 14 bits resolution.

---

[3] https://www.ti.com/product/ADS1298
[4] https://www.st.com/en/microcontrollers-microprocessors/stm32f427-437.html
[5] https://www.kionix.com/product/KX122-1037

Our default configuration was based on 4 EMG channels plus 3 channels of accelerometer data, at 8 kHz sample rate. When connected by USB we could transmit all 7 channels at 8 kHz with 24-bit resolution. Using BLE, the data was down-sampled 64x to 125 Hz and truncated to 14 bits (after pre-processing).

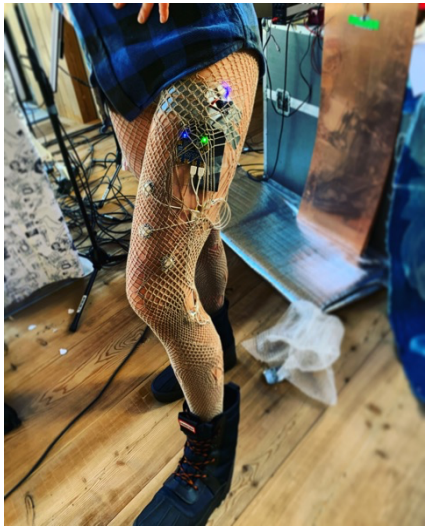

Figure 4. EAVI in performance at Muzeum Susch

### 3.4 OWL Microcontroller Framework

The OWL platform provides an abstraction layer between digital signal processing (DSP) and firmware development, without sacrificing DSP performance [11]. At least 98% of total microcontroller clock cycles can be used by the DSP process, without any audio dropouts, while firmware and interface functions are executed without delay by leveraging hardware interrupts and background DMA transfers. Between the DSP process (referred to as the patch, or user program) and the firmware there is a binary interface: the program vector. User interactions are abstracted as parameters, buttons, audio and MIDI streams. The OWL platform offers tools to develop patches in any of the following languages: C++; FAUST; Pure Data; Max gen~; SOUL, Maximilian

Patches can be compiled offline, with a Makefile system, or with an online compiler. In addition to producing the ARM binary, the tools can also compile and run, or debug, a native version. Optionally a Javascript binary can be generated and tested in a web browser.

Thanks to dynamic patch loading, the firmware doesn't have to be compiled and flashed each time the patch changes. Instead, the patch binary is packaged as MIDI SysEx, sent by USB to the device, and dynamically loaded and executed. In summary, the OWL platform allows for fast prototype, development and test cycles, with no dependency on specific hardware features.

The plans are published as open hardware at: https://github.com/RebelTechnology/BioSignals

### 4. ACKNOWLEDGMENTS
The development of this work has been supported by funding from the European Research Council (ERC Horizon2020 789825) and the French Agence Nationale de la Recherche (ANR-21-CE38-0018).